\documentclass{osa-article}

\journal{ome}
\articletype{Research Article}
\begin{document}

\title{Optical Properties of Yttrium Gallium Garnet}

\author{Jacob H. Davidson,\authormark{1} Dorian Oser,\authormark{1} and Wolfgang Tittel\authormark{1,2,3}}

\address{\authormark{1} QuTech and Kavli Institute of Nanoscience, Delft University of Technology, 2600 GA Delft, The Netherland\\
\authormark{2} Department of Applied Physics, University of Geneva, 1211 Geneva 4 \\
\authormark{3} Schaffhausen Institute of Technology -- SIT, Geneva, Switzerland\\}

\email{\authormark{*}j.h.davidson@tudelft.nl} %% email address is required



\begin{abstract}
We report measurements of the reflection and transmission spectra of 2\% doped Thulium Yttrium Gallium Garnet (Tm:YGG) using variable-angle spectroscopic ellipsometry (VASE) over a wavelength range from 210 to 1680~nm  (0.73-5.9~eV). The well-known Tm resonances are identified and separated from the aggregate data, allowing us to calculate the previously unknown frequency dependence of the complex refractive index of the host material. This information is important for many applications of YGG in classical and quantum photonics, including constructing optical cavities, laser-based applications, and quantum information devices. A complete database of the obtained  parameters is included in the supplementary information.  
\end{abstract}
%%%%%
\section{Introduction}
	
    Rare-earth-doped aluminum garnets such as neodymium- or thulium-doped yttrium aluminium garnets and lutetium aluminium garnets (Tm:YAG,Tm:LuAG, Nd:YAG) are well known for their use as laser crystals \cite{Fan1998,Ross1978,Beyatli:19}. This and other applications in classical and quantum photonics \cite{Burnett2006,Davidson2020} have been enabled by detailed spectroscopic studies, including of the crystals' complex refractive index. However, due to their great potential for many photonics technologies, investigations of additional garnets with similar mechanical and chemical properties (hardness, chemical stability, etc.) yet different optical properties are essential. 
    
    One alternative class of such materials is that of gallium based garnets, for instance gadolinium scandium gallium garnet (GSGG), which shares many of the material benefits with the aluminum garnets but feature a higher index of refraction \cite{Powell1998}. Another interesting example is yttrium gallium garnet (Y$_3$Ga$_5$O$_{12}$, YGG). Doped with various rare earth ions, it has been used to create lasers or waveguides at several frequencies \cite{Zhang2011,Yu2010}. Additionally, YGG has shown potential for creating lenses for high energy lithography \cite{Burnett2006}, LIDAR \cite{Beecher2016}, and for quantum information devices \cite{Thiel2014,Thiel20142}. With such a broad base of applications, an equally broad understanding of the material optical properties is important. However, to the best of our knowledge, only the spectral dependence of the real part of the refractive index of YGG has been reported over a broad wavelength range \cite{Enke1978} while the complex index is only known for a few isolated wavelengths \cite{ZHAO2019,Lili2019,KUROSAWA2013,Grant-Jacob2017}.

	Here we fill this gap by determining the real and imaginary parts of the refractive index of a 2\% Tm$^{3+}$:YGG single crystal between 210 and 1680 nm wavelength (0.73-5.9 eV). The well-known Tm resonances are identified and separated from the aggregate data, allowing us to calculate the previously unknown frequency dependence of the complex refractive index of the host material. Experimental data is gathered using variable angle spectroscopic ellipsometry (VASE) -- it will serve as a key resource for many applications of YGG in classical and quantum photonics.

\begin{figure*}[htbp]
\centering
\includegraphics[width=0.80\textwidth]{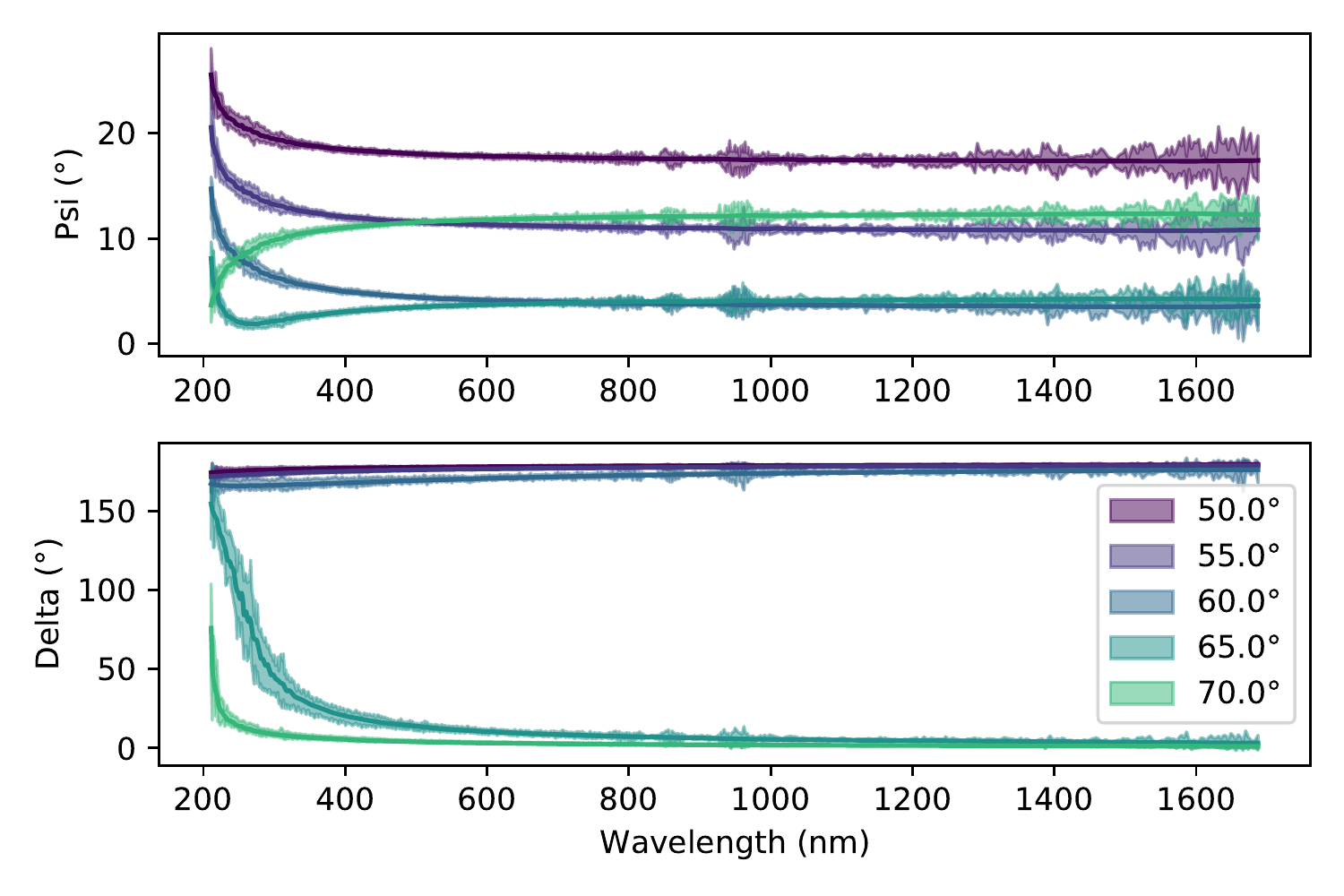}
\caption {$\Psi$ and $\Delta$ curves (see main text) for measurements at various angles of incidence. Curves shown in different colors indicate measured data for different angles. The fitted VASE model is shown by the solid lines and shows good agreement with the measured data (the mean square error is 5.06).}
\label{PsiDelta}
\end{figure*}
	
\section{Experimental Setup}
    
	  In VASE measurements, linearly polarized light with varying polarization, wavelength and incidence angle is reflected off the surface of a polished crystal in an elliptically polarized state. Analysis using a rotating polarizer allows establishing the complex reflection coefficients $R_p,\,R_s$ for the \textit{p} and \textit{s} polarized components of the reflected light \cite{Woolam1999,Blaine1999}. In turn, this allows one to calculate a pair of quantities, $\Psi$ and $\Delta$, for each input polarization, frequency, and angle. $\Psi$ and $\Delta$ are defined by
        \begin{align}
        \rho = \frac{R_p}{R_s} = \tan(\Psi)e^{i\Delta}.
        \end{align}

    The measurements were performed using a piece of single YGG crystal grown at Scientific Materials Corp. by means of their ultra high precision Czochralski growth method. The sample was doped with an additional 2\% thulium atoms, which substitute for yttrium sites in the crystal lattice. It was cut into a 5 mm by 8 mm by 10 mm rectangular prism selected from a low-strain portion of the boule. Due to the cubic symmetry of YGG, the results for reflection from every facet depend only on the angle between the incident beam and the normal to the reflecting surface  \cite{Weber2018}. All surfaces for the transmission and VASE measurements were polished to optical quality.

\section{Results}

The ellipsometry measurements were conducted on the 5 x 8 mm face using a variable-angle J.A. Woolam M 2000 Ellipsometer. Reflection measurements were taken for a series of incidence angles between 50$^{\circ}$ and 70$^{\circ}$ in steps of 5$^{\circ}$. To measure transmission we chose an incidence angle of 0$^{\circ}$---normal to the 5 x 8 mm face---for which the beam propagated along the $<1\bar{1}0>$ crystalline axis. Spectroscopic data spanning the range of 210-1680~nm (0.73-5.9~eV) was taken for all angles.

The resulting $\Psi(\lambda)$ and $\Delta(\lambda)$ curves for the measured YGG crystal are shown in Figure \ref{PsiDelta}. The figure also includes a best fit using a B-Spline function \cite{Forouhi1986,Jellison1996,Weber2009,Mohrmann2020} produced using Woolam's CompleteEase software under the assumption of an infinite substrate. The mean standard error (MSE) is 5.06, which shows good agreement between the model and the experimental data, comparable to that of other materials \cite{Hrabovsky2021,Herzinger1998}.

%    VASE measurements begin with a known polarization, wavelength, and reflection angle of the input light beam. For each of these points, data, $\rho$ is collected and used to plot a pair of quantities $\Psi$ and $\Delta$ given by. 
%\begin{align}
%\rho = \frac{R_p}{R_s} = \tan(\Psi)e^{i\Delta}
%\end{align}
 %   Here, $\rho$ is the ratio between complex reflection coefficients $R_p,R_s$ for \textit{p} and \textit{s} polarization components of the reflected light \cite{Woolam1999,Blaine1999}. Linearly polarized light incident on the sample is reflected in an elliptically polarized state which is analyzed by a rotating polarizer to produce values of $\Psi$ and $\Delta$ for each particular input of polarization, frequency, and angle \cite{Woolam1999,Blaine1999}. With $\lambda$ the wavelength of each particular measurement, $\Psi(\lambda)$ and $\Delta(\lambda)$ curves for the measured YGG material are shown in Figure \ref{PsiDelta} with a matching model for an infinite substrate simulated in Woolam's CompleteEase software.

\begin{figure}[htbp]
\centering
\includegraphics[width=0.8\textwidth]{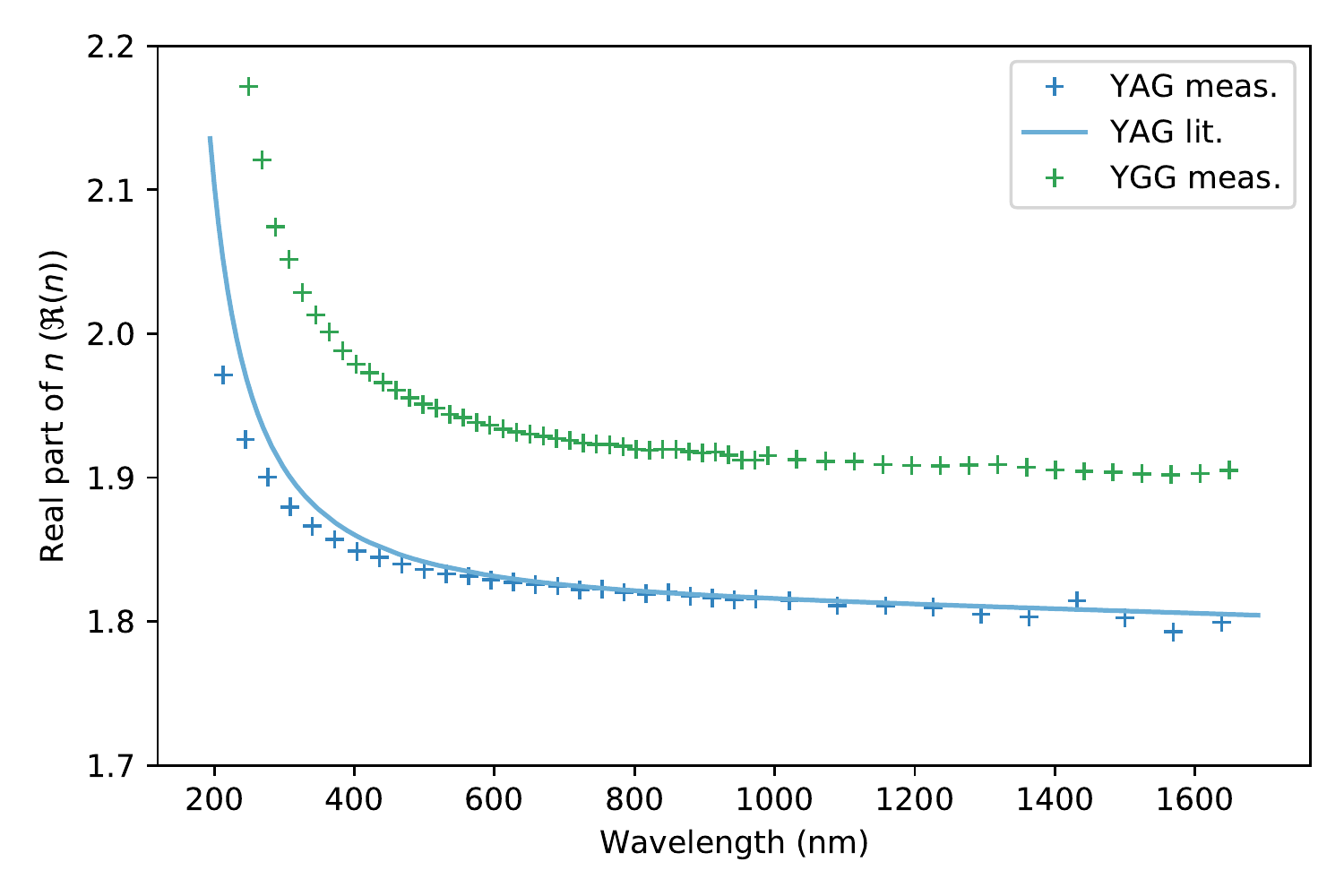}
\caption {Spectral dependence of the real part of the refractive index of YGG (green crosses). The curve is derived from the fitted data shown in figure \ref{PsiDelta}. To validate our result, we also plot the index of refraction of a 0.1\% Tm:YAG crystal, established using the same method (blue crosses), and, for comparison, the literature values (solid blue line) \cite{Hrabovsky2021}.}
\label{Index YGG}
\end{figure}

   The B-spline fit is subsequently used to calculate the wavelength-dependent real part of the index of refraction, n, across the entire measured wavelength range. Note that due to the cubic symmetry, n is isotropic. The resulting values, obtained using the ellipsometer' modeling software, is shown by the green data points in Figure \ref{Index YGG}. Furthermore, the figure includes measured data and literature values for a 0.1\% Tm:YAG crystal \cite{Hrabovsky2021}. The good agreement validates our measurement and fitting methods. Furthermore, we find that the refractive index of YGG is clearly higher than that of YAG, a result of the presence of gallium in the chemical composition of the crystal. Note that in VASE measurements, the single surface reflection results in very little interaction between the probe beam and the bulk material, leaving the measured data of Tm:YGG unaltered by the presence of a small percentage of Tm ions.

\begin{figure*}[htbp]
\centering
\includegraphics[width=0.8\textwidth]{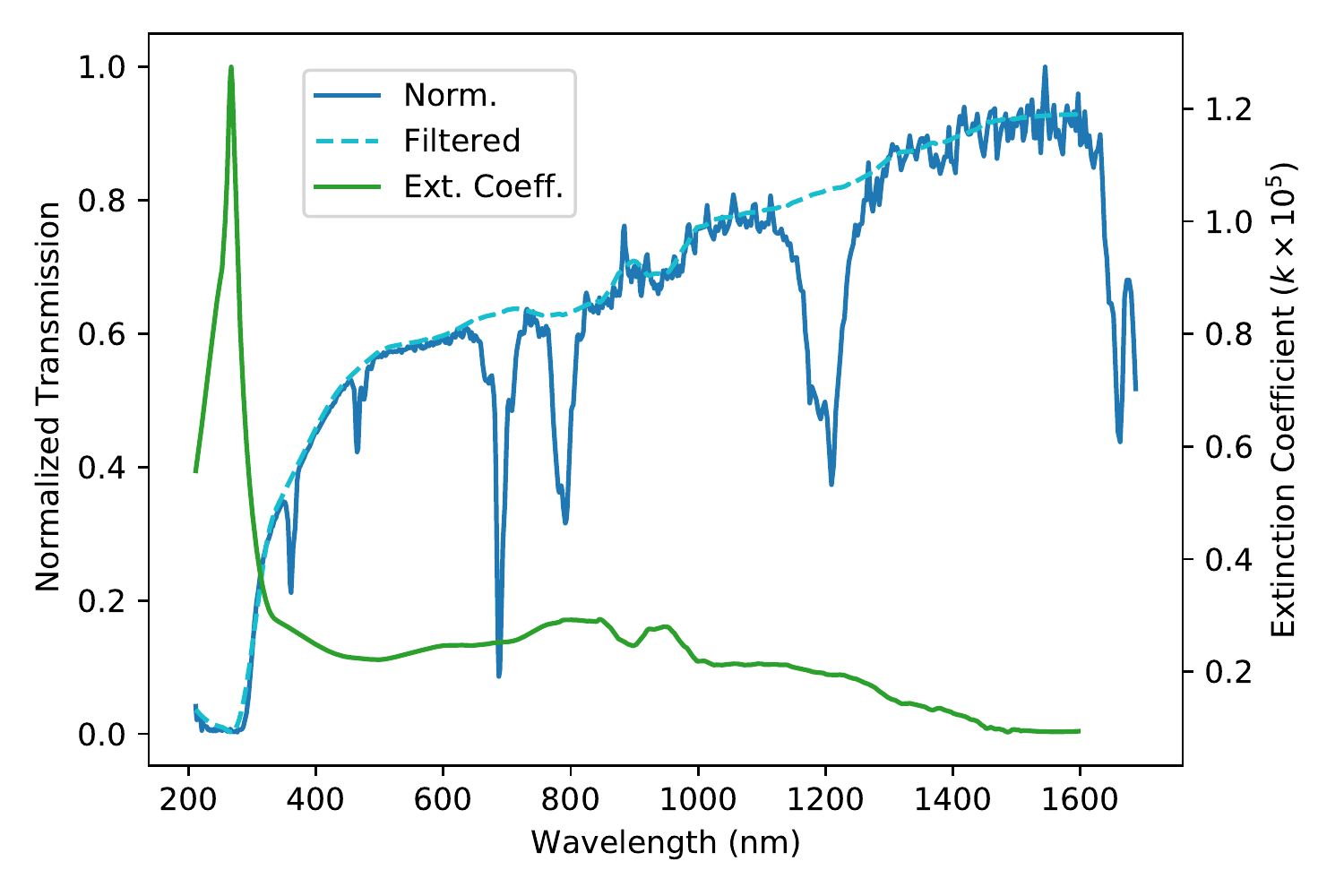}
\caption {Transmission spectrum of a 2\% Tm:YGG sample. The transmission data, in blue, clearly shows the signature of the Tm$^{3+}$ ions. All sharp dips in this spectrum can be associated with a particular atomic resonance of Tm \cite{Peterka2011,Agger2004,Volkovich2016}. The dashed curve shows the result of an interpolation that removes the affected spectral portions, and the green curve depicts the filtered extinction coefficient of YGG.}
\label{BareYGGSpectrum}
\end{figure*}

	Next, to determine the extinction coefficient, $k$, of YGG, we measure the transmission spectrum of the Tm:YGG crystal. The result is shown in figure \ref{BareYGGSpectrum}. The presence of Tm in the sample becomes immediately clear -- it manifests itself in the characteristic Tm absorption lines around 460, 680, 790, 1210, and 1600~nm \cite{Peterka2011,Agger2004}. Using Python's 1-D interpolation function, we remove (filter) these portions from the spectrum and consider the resulting transmission curve to be a property of YGG. From this normalized transmission data we calculate the extinction coefficient of the material using $k=\frac{\alpha \lambda}{4\pi}$, where $\alpha$ is the optical depth for transmission through  10 mm of Tm:YGG, and $\lambda$ is the wavelength. The wavelength dependence of the extinction coefficient  $k$ is shown in figure \ref{BareYGGSpectrum}. Values for $k$ are extremely small, showing the high optical quality of the crystal \cite{aspnes1983dielectric}.  The supplementary information contains a table of the obtained values for the complex index of refraction across all measured wavelengths.

\section{Conclusion}
	We have measured the wavelength dependence of the real and imaginary parts of the refractive index for yttrium gallium garnet over a broad spectral range between 210-1680~nm (0.73-5.9 eV). This study provides an important resources for those looking to create optical devices from YGG for a number of different applications. 

\section*{Acknowledgments}
The authors thank L. Shreik for discussions and training on the ellipsometer. We acknowledge funding through the Netherlands Organization for Scientific Research, and the European Union’s Horizon 2020 Research and Innovation Program under Grant Agreement No. 820445 and Project Name Quantum Internet Alliance.\\

% \section{Supplementary Info}

%%%%%%%%%%%%%%%%%%%%%%% References %%%%%%%%%%%%%%%%%%%%%%%%%

%%%%%%%%%% If using BibTeX:
\bibliography{biblio}

\begin{thebibliography}{10}
\newcommand{\enquote}[1]{``#1''}

\bibitem{Fan1998}
T.~Fan, G.~Huber, R.~Byer, and P.~Mitzscherlich, \enquote{Spectroscopy and
  diode laser-pumped operation of {T}m,{H}o:{YAG},} {\protect\JournalTitle{IEEE
  Journal of Quantum Electronics}} \textbf{24}, 924--933 (1988).

\bibitem{Ross1978}
M.~Ross, P.~Freedman, J.~Abernathy, G.~Matassov, J.~Wolf, and J.~Barry,
  \enquote{Space optical communications with the {N}d: {YAG} laser,}
  {\protect\JournalTitle{Proceedings of the IEEE}} \textbf{66}, 319--344
  (1978).

\bibitem{Beyatli:19}
E.~Beyatli, B.~Sumpf, G.~Erbert, and U.~Demirbas, \enquote{Efficient {T}m:{YAG}
  and {T}m:{L}u{AG} lasers pumped by 681\&\#x2009;\&\#x2009;nm tapered diodes,}
  {\protect\JournalTitle{Appl. Opt.}} \textbf{58}, 2973--2980 (2019).

\bibitem{Burnett2006}
J.~H. Burnett, S.~G. Kaplan, E.~L. Shirley, D.~Horowitz, W.~Clauss,
  A.~Grenville, and C.~V. Peski, \enquote{{High-index optical materials for
  193nm immersion lithography},} in \emph{Optical Microlithography XIX,}  vol.
  6154 D.~G. Flagello, ed., International Society for Optics and Photonics
  (SPIE, 2006), pp. 445 -- 456.

\bibitem{Davidson2020}
J.~H. Davidson, P.~Lefebvre, J.~Zhang, D.~Oblak, and W.~Tittel,
  \enquote{Improved light-matter interaction for storage of quantum states of
  light in a thulium-doped crystal cavity,} {\protect\JournalTitle{Phys. Rev.
  A}} \textbf{101}, 042333 (2020).

\bibitem{Powell1998}
R.~Powell, \emph{Physics of Solid-State Laser Materials}, Atomic, Molecular and
  Optical Physics Series (Springer New York, 1998).

\bibitem{Zhang2011}
Y.~Zhang, Z.~Wei, Q.~Wang, L.~Liang, X.~Zhong, Z.~Zhang, D.~Li, H.~Yu,
  H.~Zhang, and J.~Wang, \enquote{Diode pumped efficient continuous-wave
  {Y}b:{YGG} laser,} in \emph{CLEO: 2011 - Laser Science to Photonic
  Applications,}  (2011), pp. 1--2.

\bibitem{Yu2010}
H.~Yu, K.~Wu, B.~Yao, H.~Zhang, Z.~Wang, J.~Wang, X.~Zhang, and M.~Jiang,
  \enquote{Efficient triwavelength laser with a {N}d:{YGG} garnet crystal,}
  {\protect\JournalTitle{Opt. Lett.}} \textbf{35}, 1801--1803 (2010).

\bibitem{Beecher2016}
S.~J. Beecher, J.~A. Grant-Jacob, P.~Hua, D.~P. Shepherd, R.~W. Eason, and
  J.~I. Mackenzie, \enquote{{E}r:{YGG} planar waveguide amplifiers for lidar
  applications,} in \emph{Lasers Congress 2016 (ASSL, LSC, LAC),}  (Optical
  Society of America, 2016), p. JTh2A.9.

\bibitem{Thiel2014}
C.~W. Thiel, N.~Sinclair, W.~Tittel, and R.~L. Cone,
  \enquote{{T}m$^{3+}$:{Y}$_{3}${G}a$_{5}${O}$_{12}$ materials for spectrally
  multiplexed quantum memories,} {\protect\JournalTitle{Phys. Rev. Lett.}}
  \textbf{113}, 160501 (2014).

\bibitem{Thiel20142}
C.~W. Thiel, N.~Sinclair, W.~Tittel, and R.~L. Cone, \enquote{Optical
  decoherence studies of {T}m$^{3+}$:{Y}$_{3}${G}a$_{5}${O}$_{12}$,}
  {\protect\JournalTitle{Phys. Rev. B}} \textbf{90}, 214301 (2014).

\bibitem{Enke1978}
K.~Enke and W.~Tolksdorf, \enquote{Continuously recording refractive index
  spectrograph for transparent and opaque insulators and semiconductors,}
  {\protect\JournalTitle{Review of Scientific Instruments}} \textbf{49},
  1625--1628 (1978).

\bibitem{ZHAO2019}
J.~Zhao, L.~Ye, Y.~Liu, S.~Li, G.~Fu, and Q.~Yue, \enquote{Optical properties
  and surface blistering visualization on multiple-energy {H}e-implanted
  {Y}b:{YGG} crystal by annealing treatment,} {\protect\JournalTitle{Results in
  Physics}} \textbf{15}, 102621 (2019).

\bibitem{Lili2019}
Y.~Lili, L.~Yong, L.~Shuang, F.~Gang, Y.~Qingyang, and Z.~Jinhua,
  \enquote{Refractive index enhanced well-type waveguide in {N}d:{YGG} crystal
  fabricated by swift {K}r$^{+8}$-ion irradiation,} {\protect\JournalTitle{Opt.
  Mater. Express}} \textbf{9}, 1907--1914 (2019).

\bibitem{KUROSAWA2013}
S.~Kurosawa, V.~V. Kochurikhin, A.~Yamaji, Y.~Yokota, H.~Kubo, T.~Tanimori, and
  A.~Yoshikawa, \enquote{Development of a single crystal with a high index of
  refraction,} {\protect\JournalTitle{Nuclear Instruments and Methods in
  Physics Research Section A: Accelerators, Spectrometers, Detectors and
  Associated Equipment}} \textbf{732}, 599--602 (2013). Vienna Conference on
  Instrumentation 2013.

\bibitem{Grant-Jacob2017}
J.~A. Grant-Jacob, S.~J. Beecher, H.~Riris, A.~W. Yu, D.~P. Shepherd, R.~W.
  Eason, and J.~I. Mackenzie, \enquote{Dynamic control of refractive index
  during pulsed-laser-deposited waveguide growth,} {\protect\JournalTitle{Opt.
  Mater. Express}} \textbf{7}, 4073--4081 (2017).

\bibitem{Woolam1999}
J.~A. Woollam, B.~D. Johs, C.~M. Herzinger, J.~N. Hilfiker, R.~A. Synowicki,
  and C.~L. Bungay, \enquote{{Overview of variable-angle spectroscopic
  ellipsometry (VASE): I. Basic theory and typical applications},} in
  \emph{Optical Metrology: A Critical Review,}  vol. 10294 G.~A. Al-Jumaily,
  ed., International Society for Optics and Photonics (SPIE, 1999), pp. 3 --
  28.

\bibitem{Blaine1999}
B.~Johs, J.~A. Woollam, C.~M. Herzinger, J.~N. Hilfiker, R.~A. Synowicki, and
  C.~L. Bungay, \enquote{{Overview of variable-angle spectroscopic ellipsometry
  (VASE): II. Advanced applications},} in \emph{Optical Metrology: A Critical
  Review,}  vol. 10294 G.~A. Al-Jumaily, ed., International Society for Optics
  and Photonics (SPIE, 1999), pp. 58 -- 87.

\bibitem{Weber2018}
M.~Weber, \emph{Handbook of Optical Materials}, Laser \& Optical Science \&
  Technology (CRC Press, 2018).

\bibitem{Forouhi1986}
A.~R. Forouhi and I.~Bloomer, \enquote{Optical dispersion relations for
  amorphous semiconductors and amorphous dielectrics,}
  {\protect\JournalTitle{Phys. Rev. B}} \textbf{34}, 7018--7026 (1986).

\bibitem{Jellison1996}
G.~E. Jellison and F.~A. Modine, \enquote{Parameterization of the optical
  functions of amorphous materials in the interband region,}
  {\protect\JournalTitle{Applied Physics Letters}} \textbf{69}, 371--373
  (1996).

\bibitem{Weber2009}
J.~W. Weber, T.~A.~R. Hansen, M.~C.~M. van~de Sanden, and R.~Engeln,
  \enquote{B-spline parametrization of the dielectric function applied to
  spectroscopic ellipsometry on amorphous carbon,}
  {\protect\JournalTitle{Journal of Applied Physics}} \textbf{106}, 123503
  (2009).

\bibitem{Mohrmann2020}
J.~Mohrmann, T.~E. Tiwald, J.~S. Hale, J.~N. Hilfiker, and A.~C. Martin,
  \enquote{Application of a b-spline model dielectric function to infrared
  spectroscopic ellipsometry data analysis,} {\protect\JournalTitle{Journal of
  Vacuum Science \& Technology B}} \textbf{38}, 014001 (2020).

\bibitem{Hrabovsky2021}
J.~Hrabovsk\'{y}, M.~Ku\v{c}era, L.~Palou\v{s}ov\'{a}, L.~Bi, and M.~Veis,
  \enquote{Optical characterization of {Y}$_3${A}l$_5${O}$_{12}$ and
  {L}u$_3${A}l$_5${O}$_{12}$ single crystals,} {\protect\JournalTitle{Opt.
  Mater. Express}} \textbf{11}, 1218--1223 (2021).

\bibitem{Herzinger1998}
C.~M. Herzinger, B.~Johs, W.~A. McGahan, J.~A. Woollam, and W.~Paulson,
  \enquote{Ellipsometric determination of optical constants for silicon and
  thermally grown silicon dioxide via a multi-sample, multi-wavelength,
  multi-angle investigation,} {\protect\JournalTitle{Journal of Applied
  Physics}} \textbf{83}, 3323--3336 (1998).

\bibitem{Peterka2011}
P.~Peterka, I.~Kasik, A.~Dhar, B.~Dussardier, and W.~Blanc,
  \enquote{Theoretical modeling of fiber laser at 810 nm based on thulium-doped
  silica fibers with enhanced 3h4 level lifetime,} {\protect\JournalTitle{Opt.
  Express}} \textbf{19}, 2773--2781 (2011).

\bibitem{Agger2004}
S.~Agger, J.~H. Povlsen, and P.~Varming, \enquote{Single-frequency
  thulium-doped distributed-feedback fiber laser,} {\protect\JournalTitle{Opt.
  Lett.}} \textbf{29}, 1503--1505 (2004).

\bibitem{Volkovich2016}
V.~A. Volkovich, A.~B. Ivanov, S.~M. Yakimov, D.~V. Tsarevskii, O.~A.
  Golovanova, V.~V. Sukhikh, and T.~R. Griffiths, \enquote{Electronic
  absorption spectra of rare earth (iii) species in {N}a{C}l–2{C}s{C}l
  eutectic based melts,} {\protect\JournalTitle{AIP Conference Proceedings}}
  \textbf{1767}, 020023 (2016).

\bibitem{aspnes1983dielectric}
D.~E. Aspnes and A.~Studna, \enquote{Dielectric functions and optical
  parameters of {S}i, {G}e, {G}a{P}, {G}a{A}s, {G}a{S}b, {I}n{P}, {I}n{A}s, and
  {I}n{S}b from 1.5 to 6.0 ev,} {\protect\JournalTitle{Physical review B}}
  \textbf{27}, 985 (1983).

\end{thebibliography}

%%%%%%%%%% If preparing manually:
% \newcommand{\enquote}[1]{``#1''}

% \bibitem{Zhang:14}
% Y.~Zhang, S.~Qiao, L.~Sun, Q.~W. Shi, W.~Huang, L.~Li, and Z.~Yang,
%   \enquote{Photoinduced active terahertz metamaterials with nanostructured
%   vanadium dioxide film deposited by sol-gel method,}
%   {\protect\JournalTitle{Optics Express}} \textbf{22}, 11070--11078 (2014).

% {Optical Society}, \enquote{{OSA Publishing},}
%   \url{http://www.osapublishing.org}.

% \bibitem{FORSTER2007}
% P.~Forster, V.~Ramaswamy, P.~Artaxo, T.~Bernsten, R.~Betts, D.~Fahey,
%   J.~Haywood, J.~Lean, D.~Lowe, G.~Myhre, J.~Nganga, R.~Prinn, G.~Raga,
%   M.~Schulz, and R.~V. Dorland, \enquote{Changes in atmospheric consituents and
%   in radiative forcing,} in \enquote{Climate Change 2007: The Physical Science
%   Basis. Contribution of Working Group 1 to the Fourth assesment report of
%   Intergovernmental Panel on Climate Change,}  S.~Solomon, D.~Qin, M.~Manning,
%   Z.~Chen, M.~Marquis, K.~B. Averyt, M.~Tignor, and H.~L. Miler, eds.
%   (Cambridge University Press, 2007).

% \end{thebibliography}

\end{document}